\documentclass[aps,pre,reprint,doublespace]{revtex4-1}%
\usepackage{amsmath}%
\usepackage{amsfonts}%
\usepackage{amssymb}%
\usepackage{graphicx}
\usepackage{epstopdf}
\usepackage{natbib}
\usepackage{ulem}
\usepackage[colorlinks]{hyperref}%

\begin{document}
\title{Distribution of blackouts in the power grid and the Motter and Lai model.}
\author{Yosef Kornbluth$^{1}$, Gabriel Cwilich$^1$, and Sergey V. Buldyrev$^1$,
 Saleh Soltan$^{2,3}$, and Gil Zusman$^4$\\
$^1$Department of Physics, Yeshiva University, 500 West 185th Street, New
York, New York 10033\\
$^2$Princeton University, Princeton, NJ, USA\\
$^3$Amazon.com, New York, NY, USA\\
$^4$Department of Electrical Engineering, Columbia University, New York, NY, USA\\
}
\begin{abstract}
  Carreras, Dobson and colleagues have
  studied empirical data on the sizes of the blackouts in real grids
  and modeled them by computer simulations using the direct current
  approximation. They have found that the resulting blackout sizes are
  distributed as a power law and suggested that this is because the
  grids are driven to the self-organized critical state. In contrast,
  more recent studies found that the distribution of
  cascades is bimodal as in a first order phase transition,
  resulting in either a very small blackout or a very large blackout,
  engulfing a finite fraction of the system. Here we reconcile the two
  approaches and investigate how the distribution of the blackouts changes
  with model parameters,  including the tolerance criteria and the dynamic rules of
  failure of the overloaded lines during the cascade. In addition, we
  study the same problem for the Motter and Lai model and find
  similar results, suggesting that the physical laws of flow on the network
  are not as important as network topology, overload conditions, and
  dynamic rules of failure.
  
\end{abstract}
\date{\today}
\maketitle

\section{Introduction}
Cascading failures in the power grids continue to happen in spite of
efforts to make power grids more resilient~\cite{2003,2012,India}. The
standard criterion of resiliency is the $N-1$ criterion~\cite{Ren2008}:
the grid must safely operate in the event of the failure of any single
line. Carreras et al. have studied empirical data on the sizes
of the blackouts in real grids~\cite{Carreras2016} and modeled them by
computer simulations using the direct current (DC)
approximation~\cite{Dobson,Dobson2,Dobson3}. They have found that the resulting
blackout sizes are distributed as a power law and suggested that this
is because the grids are driven to the self-organized critical (SOC)
state~\cite{Flyvbjerg,Bak,Paczuski}. In their model, they assume that at any stage of the
cascade, one of the lines with loads exceeding the maximum values
imposed by the $N-1$ condition fails and immediately all the currents in
the grids are redistributed adjusting to the new network topology. The
motivation for this ``one-by-one'' failure rule of the cascade
propagation comes from investigation of real blackouts. It is
documented~\cite{2003,Albert} that the failures of overloaded
lines do not happen instantaneously but require a certain period of
time, during which overloaded lines undergo heating expansion. When
the expanded line touches the ground or foliage, the current in the
line dramatically increases and the line breaks, after which the current
in other lines changes almost instantaneously, so that if the current
in a previously overloaded line reduces to normal, that line's length reduces
and it may eventually survive the cascade of failures.
Ren {\it et al.}~\cite{Ren2008} suggested that the power grid is driven to the SOC state by recursive upgrading of the power grid with constantly growing power demand, applying the $N-1$ condition, or by upgrading lines involved
in the recent cascade of failures. They simulated this model of self-organization and found that the system converges in the infinite time limit to a steady state caracterized by an exponential distribution of the blackout sizes, while the
SOC models typically display power law distribution of avalanche lengths associated
with a second-order critical point as in percolation theory~\cite{Stauffer, Citinay, SoltanNEW}.

In contrast, more recent studies~\cite{spiewak,Scala} have suggested
that the distribution of cascades is bimodal, with a first order
phase transition, resulting in either a very small blackout or a very
large one.   In all these studies, the cascades of failures were
started by a random failure of a single line, and the currents were
computed from a given distribution of loads and generators using the
DC current approximation.  The difference between these studies was that the
maximal loads in ~\cite{spiewak} were computed
not by using $N-1$ criterion, but by a uniform tolerance algorithm as in
~\cite{motterLai,motter} and that the overloaded lines during each stage of
the cascade were eliminated all-at-once~\cite{Soltan2014, Citinay, SoltanNEW}, not one-by-one.

The
difference in the outcome between the ``one-by-one'' rule and the
``all-at-once'' rule suggests that the cascade propagation in the
overload models inherently depends on the dynamics of the cascade,
which makes these models very different from the simple topological
models of cascading failures in which the final outcome depends only
the network topology and initial failures and is the same for any
sequence of removals of the unfunctional elements~\cite{Hines}.

Here we reconcile the two approaches and show that the power law
distribution of cascades emerges for high protection level, and also in cases where the network topology
is close the percolation point.  We also show that the ``one-by-one''
removal rule significantly reduces the sizes of large blackouts and
their probabilities, replacing the bimodal distribution of blackouts
by an approximate power-law for intermediate protection levels when the
``all-at-once'' rule still leads to a bimodal distribution.  We show
that these features are held for both  the DC model of a power grid and
for the much simpler Motter and Lai model, suggesting that the exact
physical laws of flow on the network (Kirchhoff's laws vs. minimal path
rule) are not as relevant as the protection level, network topology
and the cascade dynamics rules.

\section{Power-law versus bimodal distributions}

In this section we will review the conditions under which the distribution of cascade sizes  in overload
models are power law or bimodal, and compare these overload models with the
topological models~\cite{Watts,Gleeson,Baxter2010,Baxter2011,Buldyrev2010,Parshani2011,DiMuro2016,DiMuro2017},  in which the outcome of
the cascade depends only on the initial topology of the network and
location of the initially damaged elements. Topological or overload
models can be investigated for two different classes of initial attacks. 

In the first well studied class~\cite{Buldyrev2010,Parshani2011,Kornbluth,Baxter2012,Dorogovtsev2006}, the cascade is started by a
massive attack, after which only a fraction $p$ of elements
survive.  For this class, an interesting problem is to
study the behavior of the order parameter $S(p)$, the fraction of the
surviving nodes at the end of the cascade, as a function of the initially
surviving elements $p$. Many topological and overload models
exhibit a first order phase transition at $p=p_t$, at which the order
parameter $S(p)$ exhibits a step discontinuity, or a second order
transition at $p=p_c$, at which the order parameter is continuous but
its derivative with respect to $p$ exhibits a step discontinuity. For
finite systems, near $p=p_t$ the distribution of the order parameter
becomes bimodal, and the transition point $p_c$ is defined as 
the point at which the population of the two peaks is equal. 
For the second order transition the distribution of the order
parameter is always unimodal. For some topological and overload models,
the line of the first order transitions, $p_t$, in a plane of two parameters  may transform into the line of second 
order phase transitions, $p_c$ at a certain value of the second parameter~\cite{Kornbluth,Parshani2011,Parshani2010,Baxter2012}.

In the second class of initial attacks, such as in the above mentioned discussion of power grids, the cascade is started
by the failure of a single element. 
Here, it is informative to study the
distribution of the ``blackout'' sizes, i.e., the number of nodes that
failed during the cascade and how such a probability distribution depends on the
model parameters.  This type of initial attacks may exist in two flavors.

In the first flavor, the system first undergoes a cascade of
failures caused by a massive initial attack with $p>p_t$ and the order parameter stabilizes at $S(p)$. This is analogous to the first class of initial attacks.
Afterwards, the avalanche is started by the removal of one additional element. In the well-studied topological
models, it is known~\cite{Dorogovtsev2006,Baxter2012}, that
\begin{equation}
S(p)-S(p_t+0) \sim (p-p_t)^{1/2},
\end{equation}
where $S(p_t+0)$ is the limit of $S(p)$ for $p\to p_t$ above the step
discontinuity.  Because in the topological models the order of removal
elements does not play any role, the removal of one node after the system
is stabilized at $p$ is identical to the initial removal of $pN+1$ nodes, 
where $N$ is the total number of elements in the system. In either situation, the
average avalanche size under this initial conditions is
$s=N[S(p+1/N)-S(p)] \approx \partial S(p)/\partial p\sim
(p-p_t)^{-1/2}$. Hence, the avalanche size diverges above the transition
point $p_t$, with the divergence characterized by the critical exponent
$\gamma=-1/2$. In network theory this transition is called a
hybrid transition, but in fact this behavior is completely equivalent
to the mean field behavior near the spinodal of the first order
phase transition, such as the behavior of the isothermal
compressibility near the spinodal of the gas-liquid phase
transition. This follows from the fact that the  isothermal
compressibility is equivalent to the susceptibility in the Ising model,
which in turn is equivalent to the average cluster size at the
percolation transition. The divergence takes place only for $p>p_t$,
while for $p<p_t$ the avalanches remain of finite size or do not
exist at all, since $S(p)=0$ for $p<p_t$. For $p\to p_t+0$, the
distribution of the avalanche sizes develops a power law behavior with an exponential cutoff $S^\ast$:
$P(s)\sim s^{-\tau}\exp(-S/S^\ast)$ with the mean field exponent
$\tau=3/2$~\cite{Dorogovtsev2006,Baxter2012},  and $S^\ast\sim (p-p_t)^{-1/\sigma}$
with $\sigma=-1$;  they, together with $\gamma= -1/2$, satisfy a usual percolation scaling relation $\gamma=(2-\tau)/\sigma$, but with different $\gamma$ and $\sigma$ than in the percolation theory~\cite{Stauffer,Bunde}.

\begin{figure}
\includegraphics[width=0.8\linewidth,angle=270]{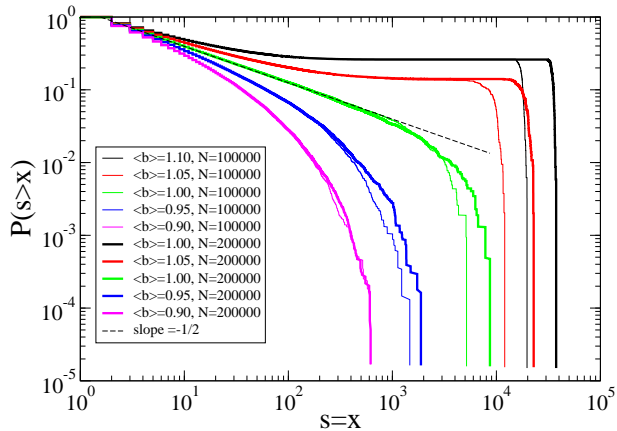}
\caption{Cumulative distributions of the avalanche sizes
  in the failure branching process with the Poisson branching degree
  distributions, with the average branching factors
  $\langle b\rangle =0.9,0.95,1,1.05,1.1$ for finite system of
  $N=100000$ and $N=200000$ nodes. One can see that the bimodal
  distribution of avalanches, characterized by a wide horizontal
  plateau of the cumulative distribution, emerges for
  $\langle b\rangle>1$. For $\langle b\rangle <1$, the distribution has
  an exponential cutoff. Exactly at the critical point
  $\langle b\rangle=1$, the distribution is a power law with a
  finite-system cutoff. The power law region with the exponent
  $\tau-1=0.5$ increases with the size of the
  system.} \label{f:branching}
\end{figure}

For the second flavor of the one-element-failure initial attack,
which is typical for most power grid simulations, the system exists in a state
characterized by some set of parameters,  but the parameter $p$ describing the size of the initial attack  is not
applicable at all. In this case, both in topological and overload models,
the cascading failures evolve as a branching process in which the
failure of one element leads to the failure of $b$ elements, which
depend on the failed element in the topological models, or get
overloaded due to the removal of the failed element in the overload
models. The critical point of the simplest branching process with a
fixed distribution of the branching factor $P(b)$ is defined by
$\langle b\rangle=1$~\cite{Harris}. For the mean-field variant of the fixed distribution
model, the distribution of the avalanche sizes is a power law with an
exponential cutoff $P(s)\sim s^{-\tau}\exp(-s/s^\ast)$, where
$s^\ast \sim (\langle b\rangle -1)^{-1/\sigma}$. Here $\sigma=1/2$ and 
$\tau=3/2$ are the critical exponents~\cite{Stauffer,Bunde}. These in turn
give the values of the two other critical exponents
$\gamma=(2-\tau)/\sigma=1$ and $\beta=(\tau-1)/\sigma=1$, where
$\gamma$ governs the divergence of the average finite avalanche size
for $\langle b\rangle \to 1$ and $\beta$ governs the probability of an
infinite avalanche (i.e., failed network) for $\langle b\rangle >1$.
These infinite avalanches, which are approximated in the finite system by finite avalanches that distintegrate the network, and whose size fluctuates around  $\mu N$, form the second peak of the bimodal distribution. The finite avalanches, which do not scale as a finite fraction of the network, form a first peak of the distributions, which is separated from the second
peak by a huge gap because there are no avalanches of
size $s$, such that  $s^\ast <<s <<\mu $ (Fig.~\ref{f:branching}). Thus, this simple branching process model predicts the existence of regimes with bimodal ($\langle b\rangle >1$) and power-law 
distributions with exponential cutoff ($\langle b \rangle \leq 1$), depending on the parameters of the model. Carreras et al. conjecture~\cite{Carreras2016} that the power grid is a SOC system which
somehow drives itself to the critical point of this branching process, at
which the distribution of blackouts is a power law. It is not clear how
this SOC process works.  It may be created by the long-term effect of
the application of the $N-1$ criterion~\cite{Ren2008} or, more likely, by some
empirical compromise between safety and price of keeping
$\langle b \rangle$ small: Electric companies want to eliminate large
catastrophic blackouts which exist for $\langle b \rangle>1$, but trying
to save money on the infrastructure, are willing to accept finite
blackouts. Obviously, this compromise is achieved for
$\langle b \rangle =1-\epsilon$. In the following sections we will check this
failure-branching process model by computer simulations. 

\section{Simulations of the power grids}
In order to verify that the $N-1$ condition in addition to the one-by-one failure rule
is responsible for the emergence of the power-law distributions of the
blackouts, we perform the DC simulations of several grid topologies
including a simplified model of the US Western Interconnect
(USWI)~\cite{Bernstein}, Learning Based Synthetic Power Grid (LBSPG)~\cite{Soltan2019,Soltan2018}, Degree and Distance attachment model (DADA)~\cite{spiewak},
and two models of
artificial topologies: a random 2-d graph with given maximal line length
(RML) and random regular 2-d graph with given degree of a node (RR).
In both RML and RR grids there are $n_p$ generator nodes which produce
power, $n_c$ loads which consume power and $n_t$ transmitter nodes,
which do not produce or consume power but redistribute the flow
between several power lines. We assume that all generators or loads
produce or consume equal power. We use the model, setting $n_p=n_c=1000$
and $n_t=8000$. All $N=n_p+n_c+n_t$ nodes are randomly placed on a
square of edge $L$ and periodic boundaries. In the RR model, each
node is connected to its $k$ nearest neighbors. In order to make sure
that each line connected to exactly $k$ nodes, we make a list of all
nearest neighbor pairs and start to select pairs from this list in 
ascending order of length, creating a link between them if both of the
nodes in the pair have less than $k$ links. At the end of the process, the
majority of nodes have $k$ links with a few exceptions which have only
$k-1$ links. In the RML model, each node is surrounded with a circle
of  radius $r=L\sqrt{\langle k\rangle/(\pi N)}$ and is connected to all the
nodes within this circle. Since the nodes are randomly distributed geographically, the degree distribution of nodes
becomes a Poisson distribution with average degree $\langle
k\rangle$. We used $k=4$ for RR and $\langle k\rangle=5.0$ for RML. If the grid
does not form a single connected cluster, we discard all the nodes
which do not belong to the largest connected cluster. 

To implement
the $N-1$ condition for all of our models, we proceed as follows: Assuming that the grid contains $n_L$ links, we obtain the current in a given 
link $ij$ after the removal of  each one of the other $(n_L -1)$  links, and we find the maximal
current $I_{ij}^\ast$ through that link $ij$ obtained over the set of those $(n_L -1)$ currents. We use this value $I_{ij}^\ast$ as the maximal
possible load that the link $ij$ may sustain. We start each simulation
with the removal of a random pair of links, $i_a j_a$ and $i_b j_b$, and
recompute the currents in all remining links, $I^1_{ij}$. If in some
links $I^1_{ij}>I_{ij}^\ast$, we find the link with maximal overload
$I^1_{ij}/I_{ij}^\ast$, remove it and find the new currents
$I^2_{ij}$ (one-by-one update rule). 
We continue this process until, after removal of $n$ links,
for all remaining links, $I^n_{ij}\leq I_{ij}^\ast$.  If at a certain
stage, the grid splits into several disconnected clusters, we
apply the power production-consumption equalization for each cluster
using the minimal production-consumption rule described in~\cite{spiewak} We compute the blackout size as a fraction of the
consumed power lost in the cascade. 

\begin{figure}
  \includegraphics[width=0.8\linewidth,angle=270]{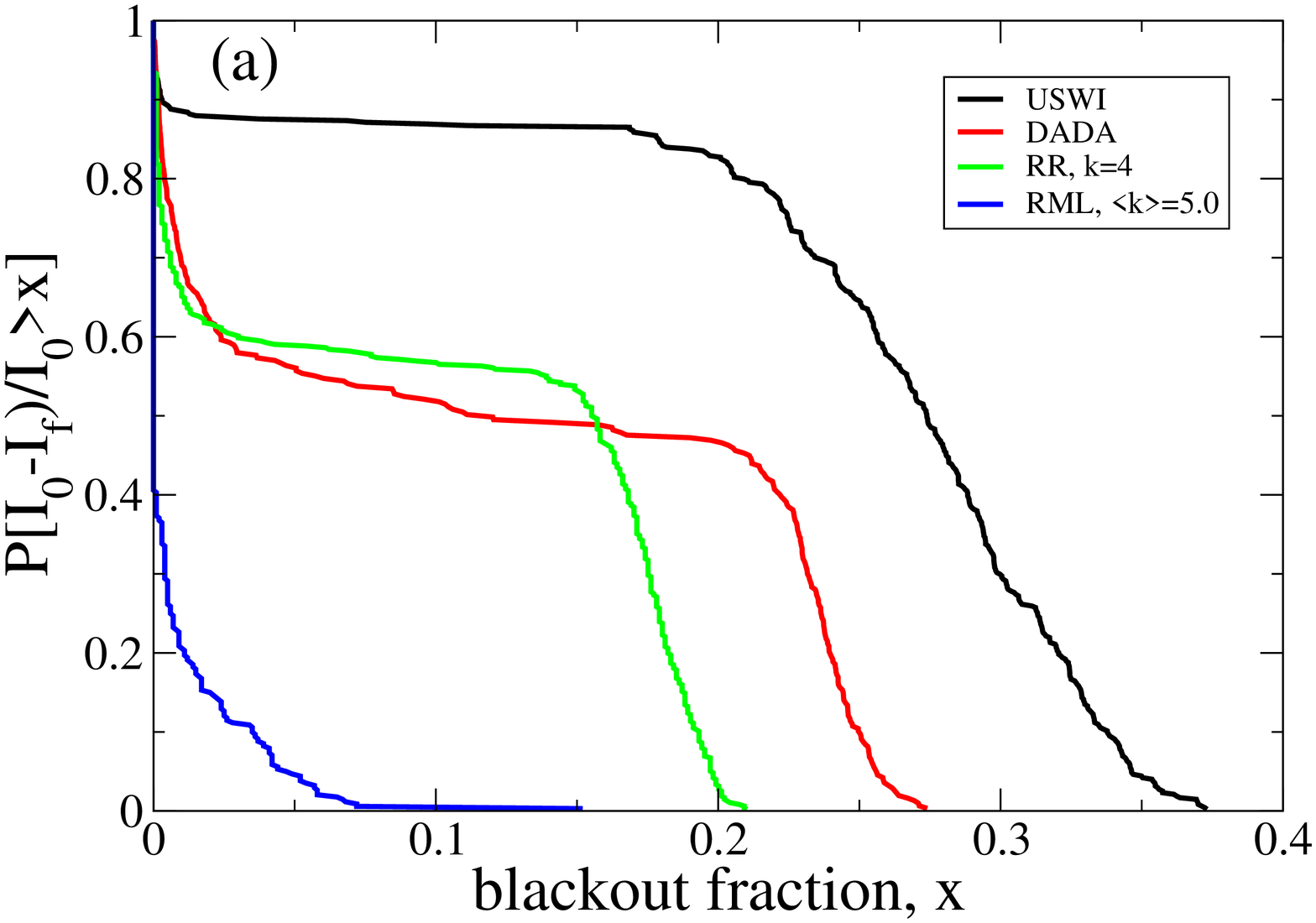}
  \includegraphics[width=0.8\linewidth,angle=0]{ln-yeild-kk.eps}
  \includegraphics[width=0.8\linewidth,angle=270]{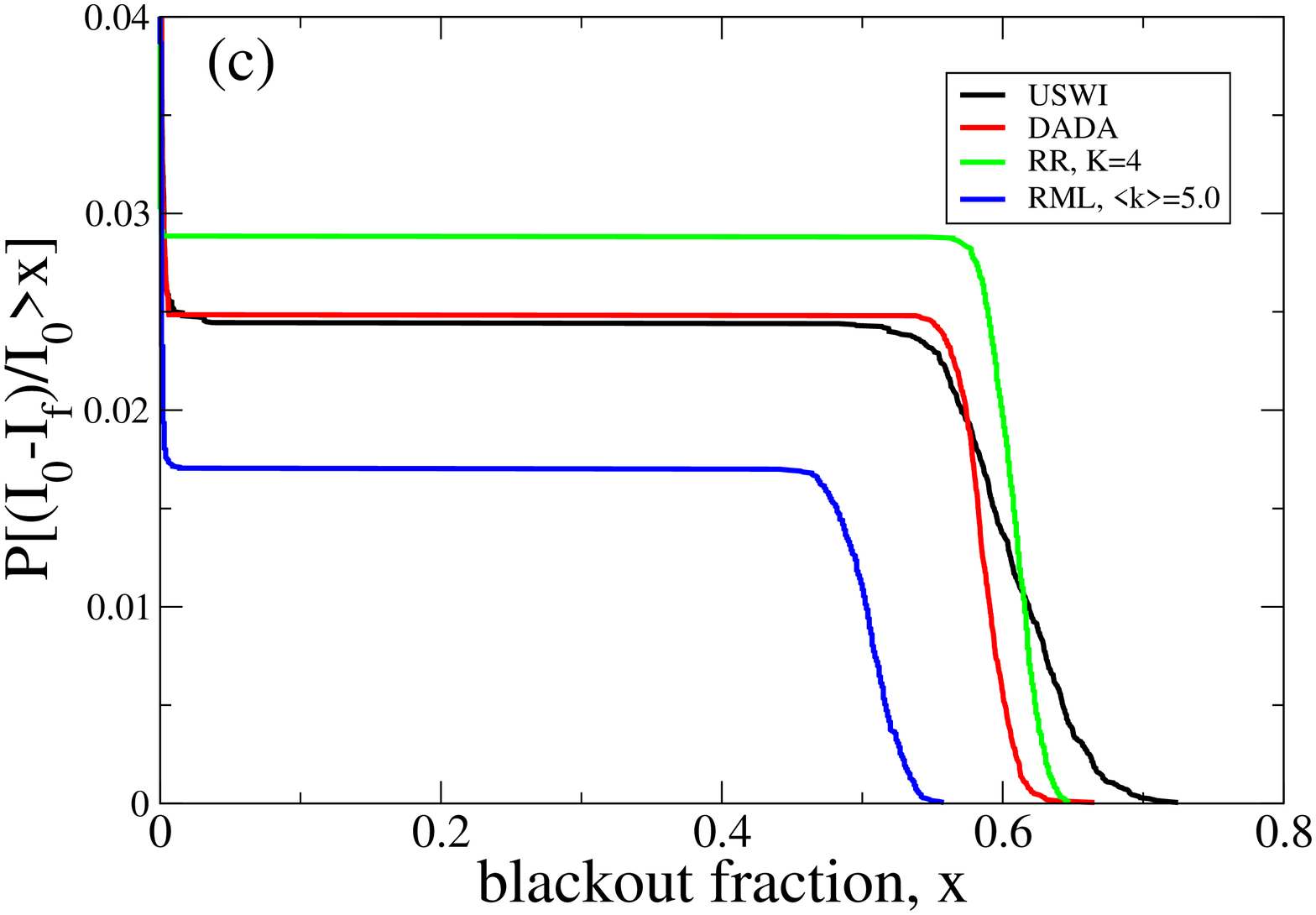}
  \caption{Cumulative distributions of blackout sizes, measured as a function of the fraction
    of the consumed power lost in the blackout for different grid topologies
    described in the text,  with the DC current approximation and the $N-1$ condition implemented.  (a) One-by-one update rule. All distributions remain bimodal except for the RML model with
    $\langle k\rangle=5.0$. (b) The same as (b) in double logarithm scale.
    A power-law distribution with $\tau-1=0.5$ emerges for the case of the RML model. (c) All-at-once update rule. All distributions remain bimodal.} \label{Fig1}
\end{figure}

In order to carefully assess the importance of the $N-1$ condition and the
update rule, we also perform simulations for networks with uniform tolerance
protection in which $I_{ij}^\ast=(\alpha+1)I_{i,j}$, where $I_{i,j}$
is the current in line $i,j$ in the unperturbed network.  Thus, for each network model, we study four
cases: $N-1$ condition together with one-by-one or all-at-once update
rules, and uniform tolerance together with one-by-one or all-at-once
update rules.\\
First, we consider the $N-1$ condition, along with the one-by-one rule. 
Implementing the $N-1$ condition significantly improves the robustness of
the grid, as compared to the uniform tolerance model with small
$\alpha$.  This is not surprising because  for all four models of the power grid, the effective
tolerance of line $ij$ under the $N-1$ rule, $\alpha_{ij} \equiv I^\ast_{ij}/I_{ij}-1$ (where
$I_{ij}$ is the initial current in the line $ij$) is a wide
distribution with almost 5\% of lines having $\alpha_{i,j}>1$. Thus, the
fraction of runs that result in any additional line overload is
about 0.03; for most pairs of initially attacked lines, no further lines are overloaded. For the tolerance model with a uniform $\alpha$, this
level of protection is achieved only for $\alpha>1$.  Thus,
implementation of the $N-1$ condition and the one-by-one rule reduces the
 sizes of blackouts and create many small cascades that end after few failures [Fig.~\ref{Fig1}(a,b)].   For all four
models of the power grid we observe the emergence of the approximate
power law part in the distribution of the size of the small cascades.  However,
for each model except RML with $\langle k \rangle=5.0$ (which is close
to the RML percolation threshold for $\langle k \rangle=4.5$), in addition to small blackouts characterized
by an approximate power law distribution, we still find some large
blackouts with a significant fraction of power loss. Note also that
the power-law part of this distribution is still practically absent in the
case for the USWI and LBSPG models. Thus, the entire distribution remains bimodal
with two peaks for small and large cascades and practically no cascades
of an intermediate size. This absence can be observed on the cumulative
distribution graph as a large plateau [Fig.~\ref{Fig1}(a)].  The value of the
exponent characterizing the power law part of the cumulative distribution
of small cascades varies for different models but remains in the range
between 0 and 1.

We next consider the $N-1$ condition, but with the all-at-once update rule.
The power law part of the blackout distribution
disappears [Fig.~\ref{Fig1}(c)]. The distribution becomes strictly bimodal
even for the case of RML model with $\langle k\rangle=5.0$, for which the $N-1$
condition together with the one-by-one rule produce a power-law distribution of blackouts. Additionally,
 large blackouts become much larger with the all-at-once rule than with the one-by-one rule; the system is more prone to failure. 

\begin{figure}
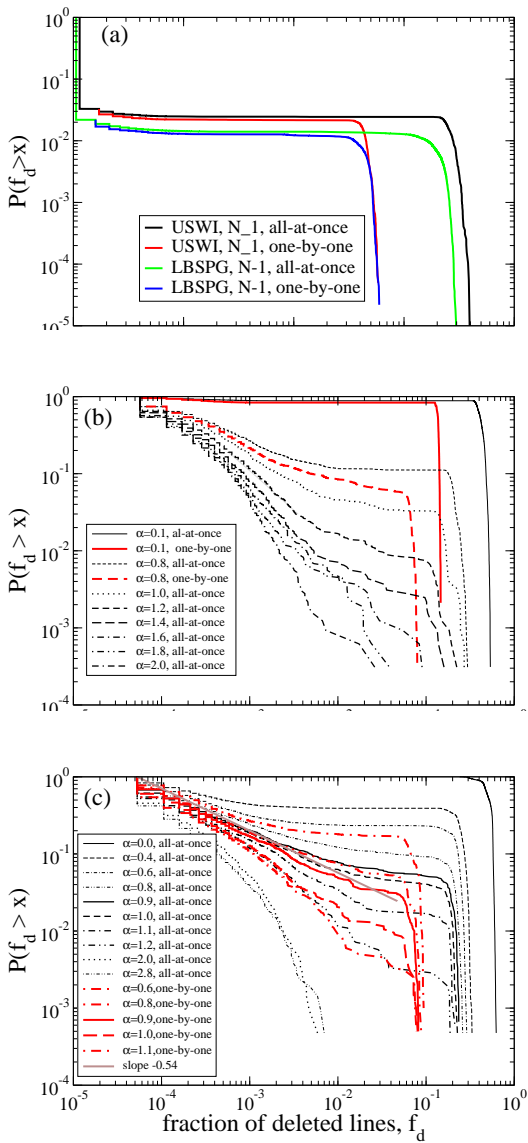

    \includegraphics[width=0.8\linewidth]{LBSPG-USWI.eps}
	
  \includegraphics[width=0.8\linewidth]{real-ln.eps}
  \\
    \includegraphics[width=0.8\linewidth,angle=0]{nk.eps}
    \caption{Cumulative distributions of blackout sizes, measured as a function of 
       the fraction of failed lines for the USWI and LBSPG models.  (a)
      Both models for the $N-1$ condition with all-at-once and one-by-one
      update rules. (b) USWI model,  and (c) LBSPG for different uniform
      tolerance $\alpha$ and both update rules. One can see that
      for sufficiently large values of the tolerance, $\alpha >1.2$, the
      distribution becomes approximately power-law with large values
      of $\tau-1 \geq 1$. The implementation of the one-by-one update rule with the
      uniform tolerance model shows that the power law distribution
      emerges for smaller value of the tolerance, $\alpha=0.8$.} \label{Fig2}
\end{figure}

Next, we systematically explore how the $N-1$ condition, uniform
tolerance, and different dynamic rules affect the blackout distributions
in the USWI and LBSPG models (Fig.~\ref{Fig2}). First of all, we see
that the
behaviors of the two models are very similar. This is not surprising because the LBSPG has been designed to reproduce the topological features of USWI. Fig.~\ref{Fig2}(a) shows that the $N-1$ condition with all-at-once rule or with one-by-one rule leads to bimodal distributions.
Although the chance to obtain a cascade of any length bigger than one is
very small,  once a cascade starts the chance of obtaining a blackout
of a significant fraction of the system size is quite large. The difference between one-by-one and all-at-once rules is that for the latter, the typical maximal blackout
corresponds to approximately 40\% of all lines, while for the former it is only 6\% of all the lines. 

If, instead, we return to the uniform tolerance condition of~\cite{spiewak}, and still use the all-at-once rule for line removal
as in~\cite{spiewak}, the shape of the blackout distribution
depends only on the tolerance level $\alpha$ [Fig.~\ref{Fig2}(b),(c)].
As in~\cite{spiewak},
we start a cascade with an initial removal of a single line. In
agreement with the original results of~\cite{spiewak}, the
bimodality disappears for $\alpha >1$ and  a power-law distribution appears instead. 
The introduction of the one-by-one update rule together
with the uniform tolerance condition shows the emergence of the power law for
lower values of the tolerance, $\alpha=0.8$, and a significant reduction of
the sizes of the largest blackouts (Fig.~\ref{Fig2}(b)).\\																						  

All in all, our simulations suggest that neither the $N-1$ condition nor
the one-by-one update rule are necessary nor sufficient for the emergence
of the power-law distribution of the blackouts. The important condition is
an overall level of line protection which can be efficiently achieved
 by increasing enough the uniform tolerance $\alpha$.  The $N-1$ condition alone does not eliminate large catastrophic blackouts in the bimodal distribution;  instead it significantly reduces the chance of any blackout caused by initial removal of two lines.
 The one-by-one update rule also
reduces the sizes of largest blackouts and removes the bimodality at
smaller values of $\alpha$,  but is not strictly necessary. Another important aspect is the network
topology. For sparse homogeneous networks whose failures are close to the
percolation threshold such as our RML model, the power-law
distribution of blackouts emerges for smaller levels of protection,
while for the other models similar failure rules still yield a bimodal distribution.

\section{Betweenness centrality overloads with the $N-1$ condition} 
As one can see, the models of the power grids we study are relatively
complex and have many different parameters, the role of which is
difficult to clarify. Therefore, it is desirable to study a simpler
model of overloads in which the role of each parameter becomes clear.
It is also interesting to see if the emergence of the power law
distribution is a universal phenomenon for different overload models.
A good candidate for such a model is the betweenness centrality model, which
displays a clear first order transition for the all-at-once removal rule
and uniform tolerance condition employed by Motter and Lai~\cite{motterLai}.
However,
we can expect that, similarly to the power grid models, the implementation of the $N-1$
condition and one-by-one removal rule in the betweenness centrality
model will also lead to the emergence of a power-law distribution of
the blackout sizes.  Testing of this hypothesis will allow us to
better understand the general mechanism of the emergence of the power
law distribution of the sizes of blackouts in the overload models.

We build the beweenness centrality model as in~\cite{Kornbluth}.
Namely, we create a randomly connected graph with a given degree
distribution $P(k)$ and compute the shortest paths between each pair
of nodes. The length of each edge is assumed to be equal to
$1+\epsilon$, where $\epsilon$ is a normally distributed random variable
with a small standard deviation $\sigma_\epsilon \ll1$. This
precaution is taken in order to make sure that each pair of nodes $ij$
has a unique shortest path connecting them. The betweenness centrality
$b(k)$ of each node $k$ is computed as the number of all shortest
paths $ij$ passing through node $k$, such that $k\neq i$ , $k\neq j$. The
$N-1$ condition in the betweenness-centrality model should be understood in the following way:  
If any node $i$ is deleted, the rest of the nodes remain
connected and the betweenness centrality of each remaining node $j$,
$b_i(j)$ does not exceed its maximal possible load $b^\ast(j)$. 
Thus
\begin{equation}
b^\ast(j)=\max_{i\neq j} b_i(j)
\end{equation}
and the original graph must be biconnected.  Note that in the original
Motter and Lai model with a uniform tolerance $\alpha>0$,
\begin{equation}
  b^\ast(j)=(1+\alpha)b(j),
\end{equation}
where $b(j)$ is the initial betweenness of node $j$ in a completely intact network. \\
To implement this model, we first construct a randomly connected graph
of $N$ nodes with a given  degree distribution $P(k)$ by the Molloy-Read
algorithm~\cite{Molloy}  and find its
largest biconnected component with $N_b$ nodes using the Hopcroft-Tarjan
algorithm~\cite{Hopcroft}.  After that, we compute the betweenness centrality of each
node and repeat $N_b$ simulations with each node removed in turn to find
$b^\ast(i)$ for each node.  To find the distribution of cascades of
failures we remove a pair of random nodes and find the nodes $i$ with
overloads $z=b(i)/b^\ast(i)>1$. In the first model of all-at-once
removal, at each stage of the cascade we remove all the
nodes with overloads, recompute the betweenness centrality of the
remaining nodes and repeat this process until no overloaded nodes are
left. In the second model of one-by-one removal, at each stage of the
cascades we remove only  one of the overloaded nodes, the one  with the largest
overload $z$,  we then recompute all the centralities, and repeat this process
until no overloaded nodes are left.  We study several cases of Erd\"os
R\'enyi (ER) networks with different average degrees
$\langle k\rangle$ and different number of nodes in the biconnected
component $N_b$. The size of the blackout is computed as the fraction of
removed nodes at the end of the cascade.  We also  study the distribution of the number of nodes
disconnected from the giant component.

\begin{figure}
  \includegraphics[width=0.8\linewidth]{pp.eps}
    \includegraphics[width=0.8\linewidth, angle=270]{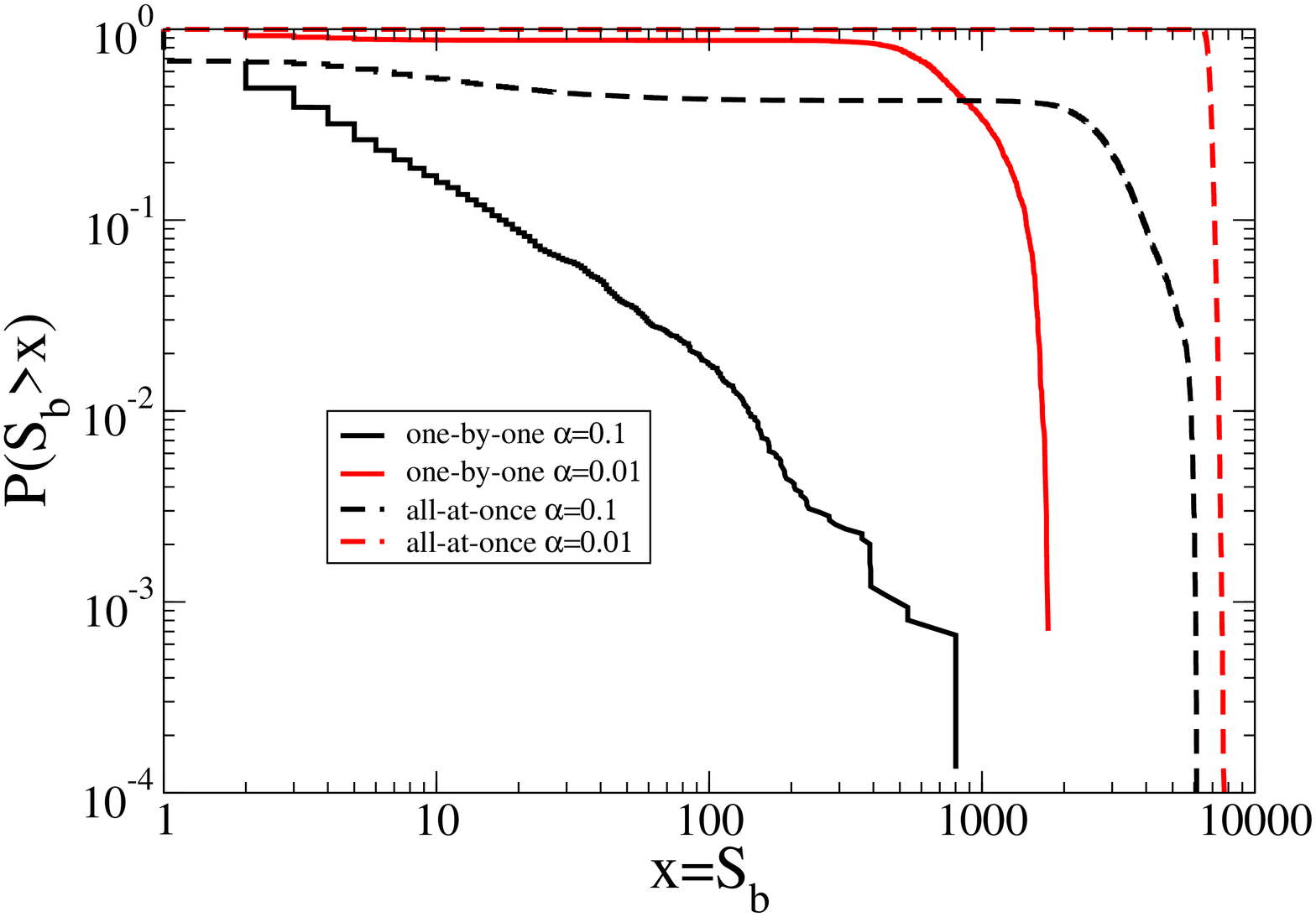}
  \caption{(a) Cumulative distributions of blackout sizes, measured as the fraction
    of failed nodes in the Motter and Lai model with different values of tolerance $\alpha$ and all-at-once update rule. Black lines indicate graphs for values of $\alpha=0.02,0.03,...0.19$. (b) Comparison of all-at-once  and one-by-one update rules for selected values of the tolerance. One can see the emergence of the power law distribution in the one-by-one removal  case for much smaller values of tolerance .} \label{Fig3}
\end{figure}

Again, the $N-1$ condition is not necessary for the emergence of the
power-law distribution of the cascades in the Motter and Lai
model. For example, if we take an ER network with
$\langle k\rangle=1.5$ and $0.1< \alpha <0.2$, we see the transition
from a bimodal distribution of the cascades to a power-law
distribution with an exponential cut-off [Fig.~\ref{Fig3}(a)] as $\alpha$ increases. 
This is somewhat similar
to the effect observed for large $\alpha$ and small fraction $p$ of
nodes that survive the initial attack in Ref.~\cite{Kornbluth}, where the
bimodal distribution of the cascades ceases to exist. In fact, small
$p$ effectively brings the network to the percolation transition
equivalent to small $\langle k\rangle \to 1$.
Replacing the all-at-once update rule by the one-by-one
rule shifts the transition from  bimodal to a power-law
distribution at a smaller $\alpha$ [Fig.~\ref{Fig3}(b)], but both behaviors can still be seen.

\begin{figure}
  \includegraphics[width=0.8\linewidth,angle=270]{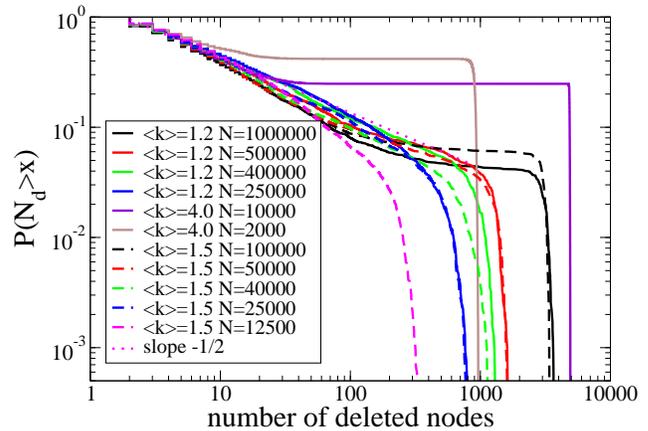}
  \caption{Cumulative distributions of blackout sizes, measured as the fraction
    of failed nodes in the Motter and Lai model with $N-1$ conditions and one-by one update rule for different system sizes and different average degrees.}
  \label{Fig4}
\end{figure}

Simultaneous application of the $N-1$ condition and the one-by-one removal
rule leads to the emergence of the power-law distribution of the small
cascades for small values of $\langle k \rangle$ (Fig.~\ref{Fig4}).
For small $\langle k \rangle$, close to the percolation
threshold and small system sizes, the bimodality of the blackout
distribution disappears and the distribution of blackouts becomes an
approximate power law with an exponential cutoff. As the size of the
network increases, the exponential cutoff disappears and the
bimodality emerges again, with a peak for large blackouts
emerging. The beginning of the cumulative distribution for all values of
$\langle k \rangle$ and all sizes of the system exhibits the same
slope of $-1/2$, which is the mean-field value for the SOC models~\cite{Flyvbjerg}.
For large $\langle k \rangle$, we observe only a small
fraction of blackouts distributed as a power law, while the entire
distribution remains bimodal; many networks completely disintegrate. However, the increase of the system size
leads to the relative increase of the range of the power law
behavior. 													  
As the system size increases, we
observe similar changes in the distribution of the blackouts for all
values of $\langle k\rangle$ studied (Fig.~\ref{Fig4}).  For
small system sizes, we observe a smaller initial negative slope $(>-1/2)$
and a sharp exponential cutoff. For larger system sizes the slope
increases to almost exactly $-1/2$, and the cutoff shifts to larger
values. For even larger sizes, the slope of the initial part of the
cumulative distribution increases even further $(<-1/2)$ but an
inflection point appears in the distribution, indicating the start of bimodality. As the
system size continues to increase the negative initial slope continues
to grow and the plateau region of the cumulative distribution
separating large blackouts from small becomes longer and
longer.

 Interestingly, the behavior of the cumulative distribution of
the blackouts for fixed $\langle k\rangle$ and increasing system size
is analogous to the behavior of the cumulative distribution of the
sizes of avalanches in the failure-branching process on a system of 
fixed size when the branching factor $\langle b \rangle$ increases (Fig.~\ref{f:branching}). The cumulative
distribution for  the avalanche distribution  for $\langle b \rangle <1 $ has a negative slope smaller than $\tau-1$,   and has a strong exponential cutoff. As $\langle b \rangle $ increases,
the slope increases and the exponential cutoff shifts to larger
values. When $\langle b\rangle$ approaches the critical point, $\langle b\rangle =1$,
the distribution
develops an inflection point and a plateau that separates finite clusters
and the giant component. For $\langle b\rangle>1$, the initial slope becomes larger
than $\tau$.  In contrast to this, the behavior of the blackout size distribution, for fixed size, as $\langle k\rangle$ approaches the
critical value of the percolation transition is
analogous to increase of the system size in the distribution of
avalanches in the failure-branching process. As $\langle k\rangle \to 1$, the range over which 
we see the power behavior with the mean field critical exponent
$\tau=-0.5$ increases. Similarly, in a failure-branching process, the length of the failure avalanche increases,
 even as its behavior and exponent do not change.

\begin{figure}
  \includegraphics[width=0.8\linewidth,angle=270]{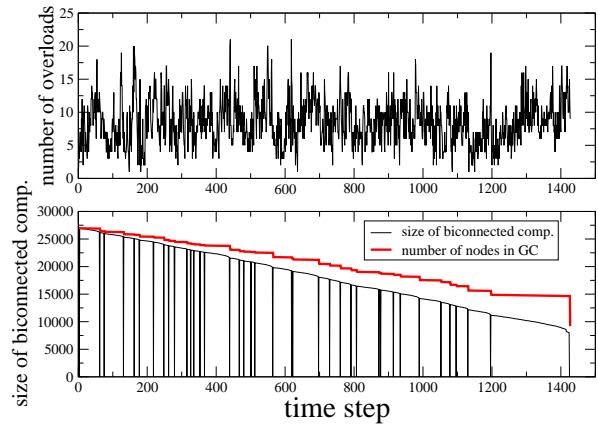}
  \caption{(a) Dependence of the number of overloaded nodes as function of the cascade step for the Motter and Lai model for $\langle k\rangle=1.2$ and $ N=500000$, with the $N-1$ condition and one-by-one update rule. (b) Dependence of the size of the biconnected component and the giant component as a function of the cascade step.}
\label{Fig5}
\end{figure}

As we saw, Carreras et al.~\cite{Carreras2016} suggested that the power law distribution of the blackouts is due to the long-term evolution of the power-grids,
driven by simultaneous growth of power consumption and upgrades of the most vulnerable power lines~\cite{Ren2008}.  
However the propagation of cascades in a power grid of fixed size, when applying the $N-1$ condition and the one-by-one update rule,  resembles SOC models, in which the most vulnerable element is removed, and the vulnerability of other elements changes instantaneously.  To test this hypothesis, we perform a
detailed study of the propagation of the cascade in a biconnected
component with $N-1$ protection and one-by-one update rule. One can see
that during the cascade the number of overloaded nodes fluctuate near
zero and never gets too large. In the mean-field variant of SOC, the
simplest of the SOC models~\cite{Flyvbjerg}, an array of $N$ real numbers,``fitnesses'', uniformly distributed between 0 and 1 is created. At each time step
the smallest element in the array and $m$ randomly selected elements
are replaced by new values selected from the same uniform
distribution. As a result, after certain transition period, the
majority of elements have fitness above the critical value
$f_c=1/(m+1)$. The number of active elements (elements with fitness
less than a certain value $f$) performs a random walk, and an avalanche
ends when the number of active elements returns to zero. When the value
of the selected fitness is equal to the critical value $f=f_c$, the random
walk is unbiased,  and the distribution of avalanche lengths coincides with the
distribution of returns to the origin of a random walker, which has the
exponent $\tau-1=1/2$~\cite{Feller}.  The number of active elements for
$f=f_c$ scales as $N^{d_f}$ with $d_f=1/2$. We see
[Fig.~\ref{Fig5}(a)] that in our model the behavior of the number of
overloaded nodes is similar to the behavior of the random walk, which
is consistent with the observed value of $\tau \approx 0.5$. However,
the DFA analysis~\cite{Peng94,Buldyrev95} of the number of overloaded nodes during the cascade
gives the value of the Hurst exponent $\approx 0.1$, which is much
smaller than the exponent for the uncorrelated random walk of the mean-field SOC
model. The number of overloaded nodes scales as $\ln N_b$, or a very
small power law. In our model, the fitness of a node is its relative
overload $z_i$, and the level of $z_i$ is not arbitrary as in the SOC
models,  but is defined initially as $z_i=1$. When, after implementing the $N-1$
condition, we initially remove two nodes, the system is placed near
the critical point, because the number of overloaded nodes after the initial attack is usually
very small.  However, this is done not automatically, as in the SOC
models, but by implementing the $N-1$ condition and the initial removal of two nodes.
This action results in the overload of  very few nodes, because the combined effect of the
removal of two nodes rarely creates overloads larger than the maximal
overloads created by the separate removal of each of these nodes. This is due to a long-tail effect;  the removal of a single node will greatly increase the load of a few nodes, but barely affect most of the other nodes.
However, when more nodes are removed in the process of the cascade,
the topology of the network changes and the system is slowly driven
away from the critical point. 

We also see that the disintegration of
the biconnected component is approximately a random removal of nodes
due to subsequent overloads, with slight preference for the removal of
nodes with $k=2$ over nodes with $k=3$.  At the beginning of the
cascade the removals take place in the biconnected component, reducing
it by the length of a chain of nodes with $k=2$ in between a pair of
nodes with $k\geq 3$. Each such removal only reduces the size of the giant
component by one. But as more and more nodes are removed, further removals may
happen in the singly connected part of the network. Figure \ref{Fig5}
(b) shows the size of the biconnected component to which the removed
node belongs as function of the cascade stage. If it belongs to a
singly connected part of the remaining network, we plot zero instead of
the size of the biconnected component. When a node that is part of the singly connected part fails, the network separates into disconnected parts and the size of the giant
component drops significantly. At the beginning, the smaller part is
usually a dangling end and the drop in size is relatively small, but eventually,  two approximately equal parts get  separated.  At this point the betweenness of all nodes
 decreases dramatically (approximately by factor of 4) and the cascade
ends.  Thus, although the cascade dynamics has some similarities with
SOC, (at each time step the node with the largest overload is removed
and the overloads of the rest of the nodes are changed), our model does
not stay at the critical state but moves away from it.

The comparison between a mean-field variant of SOC and  our network overload dynamics is also instructive to help us
understand when a bimodal distribution occurs and when a power law distribution does. 
A power law distribution, such as in a branching process, occurs when there is a series of failures, each of which
has a chance of precipitating $b=1$ additional  failures. Since there is the possibility, at each stage of the cascade,
of no new nodes failing and the process stopping, the chance that exactly one more node will fail decreases exponentially.
At the critical point, the cascade begins with $b=1$, but as the network
disintegrates $b$ changes. The  all-at-once update rule causes the network to disintegrate in fewer steps, since multiple
nodes are removed  at each step, and thus $b$ increases during the cascade of failures without many chances for the cascade to end prematurely. Thus, after a few steps $b>1$,  and the
process will not stop itself with an intermediate level of destruction. However, there are two different conditions that will prevent $b$ from
increasing drastically as the network begins to disintegrate. One is a high level of overall protection, whether through the implementation of the
$N-1$ condition or through a high uniform $\alpha$. High overall protection means that minor network destruction will not cause
widespread overload and drive $b$ away from one; only severe network destruction can do that. Thus, the overload process has a chance to stop spontaneously before it reaches that
stage. The other conditions is  if the network is near the percolation point, where the failure of a single node has a high chance of 
separating a small, but sizable component from the network. This will drastically decrease the overall load and thus
stabilize the network, arresting the cascade of failures~\cite{Kornbluth}.  In cnclusion, a one-by-one update rule and a high overall protection both contribute to the formation of a power law. 
The effect of the system size discussed above can also be understood in light of the comparison to a branching process. The larger the system,
the more nodes (as an absolute number, not a fraction) fail following the initial attack. Thus, the chance that none of the overloaded
nodes will cause another node to overload (i.e. a spontaneous end to the cascade) becomes increasingly unlikely as the number of overloaded nodes increases, even if $b$ remains close to one for many steps of the cascade.
Thus, the chance that the cascade of failures will end with an intermediate amount of damage becomes more unlikely when the size of the original system increases.

\section{Discussion and Conclusion} 

We have shown that the DC model of the power grid has remarkable
similarities to the Motter and Lai model of betweenness centrality,
suggesting that the exact physical laws governing the flux do not play
as big a role as the network topology and the dynamics of the cascade propagation.
We have shown that both overload models have similarities and
differences with the topological models with dependency
links~\cite{Parshani2011}. In the topological models the outcome of the
cascade depends only on the topology of the network and the location
of the initial damage, while in the overload models the outcome
significantly depends on the dynamics of the cascade, which itself depends on
the order of removal of overloaded elements and the relative speed of
failure of overloaded elements and redistribution of flux after
elements failure. In general, if the failure is slow compared to the
redistribution of flux (one-by-one removal rule) the cascades become
smaller and the distribution of blackouts becomes unimodal, while for
fast failures of overloaded elements (the all-at-once removal rule)
the distribution of blackouts is bimodal.  The shape of the
distribution of the blackout sizes becomes approximately a power law as the
topology of the network approaches the percolation transition for
large level of protection. In summary, the general picture of the
behavior of the overload models is consistent with the simple model of
the failure branching process. For large protection levels (small
$\langle b\rangle$) the distribution of blackout sizes is a power law
with exponential cutoff, while for low protection levels (large
$\langle b\rangle$) the distribution of blackout sizes is bimodal, with a
fraction of large avalanches which decreases as the protection level
increases.

The empirical observation of the power law distribution of the
blackout sizes in real power grids indicates that these grids are
near the critical point of the correspondent failure branching
process.  However, this is unlikely to be caused by a SOC mechanism in which
the model is driven to the critical point by a certain dynamic
rule. One hypothesis would be that it is the  $N-1$ criterion and the one-by-one
update rule that brings the grid to the SOC state. However, we observe here that
these two conditions are neither necessary nor sufficient for the existence of a power law
distribution of blackout sizes. Most likely the power grids are brought to
the vicinity of the critical point by a risk-price compromise, in which
the utilities, with the firm goal of avoiding the catastrophic blackouts that
engulf a significant fraction of the entire system,  increase
the level of protection up to the critical point of the branching
process $\langle b \rangle<1$; but in order to minimize costs under this condition, allow $\langle b \rangle$ to be as
close as possible to the critical point from below, so that limited
but occasionally very large blackouts are still possible. It is still
desirable to develop a model of long time evolution of power grids similar to~\cite{Ren2008}, which reproduces the power-law distribution of avalanches. The
observation of power law distribution of the blackouts in the sparse
networks close to the percolation point also emphasizes the role of
islanding for preventing large blackouts.

\section{Acknowledgements}
Our research was supported by HDTRA1-14-1-0017, HDTRA1-19-1-0016, and HDTRA1-13-1-0021. SVB  and GC acknowledge the partial support of this research through the Dr. Bernard W. Gamson Computational Science Center at Yeshiva College.

\end{document}